\documentclass[letterpaper, 10 pt, conference]{ieeeconf}  

\IEEEoverridecommandlockouts                              

\usepackage{cite}
\usepackage{graphicx}
\usepackage{hyperref}
\usepackage{textcomp}
\usepackage{color}
\usepackage{algorithm}
\usepackage{algpseudocode}
\usepackage{amsmath,mathtools,amssymb}
\usepackage{comment}
\usepackage{subcaption}

\newtheorem{theorem}{Theorem}

\newtheorem{proposition}[theorem]{Proposition}

\newtheorem{ass}{Assumption}
\newtheorem{definition}{Definition}

\newcommand{\R}{\mathbb{R}}

\newcommand\norm[1]{\ensuremath{\lVert#1\rVert}}

\DeclareMathOperator\diag{diag}

\DeclareMathAlphabet\mathbfcal{OMS}{cmsy}{b}{n}

\title{\LARGE \bf
Decentralised model predictive control of recurrent neural network models
}

\author{Daniele Ravasio, Marcello Farina and Andrea Ballarino
\thanks{D. Ravasio and M. Farina are with the Dipartimento di Elettronica, Informazione e Bioingegneria (DEIB), Politecnico di Milano, Italy. {\tt\small \{daniele.ravasio, marcello.farina\}@polimi.it}}%
\thanks{D. Ravasio and A. Ballarino are with the Istituto di Sistemi e Tecnologie Industriali Intelligenti per il Manifatturiero Avanzato, Consiglio Nazionale delle Ricerche, Italy. {\tt\small \{daniele.ravasio, andrea.ballarino\}@stiima.cnr.it}}%
}

\begin{document}

\maketitle
\thispagestyle{empty}
\pagestyle{empty}

\begin{abstract}
This work proposes a decentralised nonlinear model predictive control algorithm for systems described by input-coupled recurrent neural network models. The proposed approach exploits closed-loop incremental input-to-state stability to design the local controllers and provide robustness to subsystem couplings. The resulting control scheme guarantees asymptotic convergence to a desired equilibrium while satisfying input and output constraints. Simulation results illustrate the performance of the proposed approach.
\end{abstract}
%
%
%
%
\section{INTRODUCTION}
Networked systems, composed of multiple interconnected units, are complex, often large-scale and geographically distributed plants \cite{maestre2014distributed}. Representative examples include smart grids, traffic networks, and water distribution systems.\\
The application of traditional
centralised control approaches to networked systems may become impractical due to communication constraints and privacy concerns, which may prevent individual units from sharing their local measurements with a central computational unit.  Furthermore, computational issues can arise as the system size increases.\\
To overcome these limitations, a well-established approach consists of exploiting the structural modularity of the plant for decomposing the overall control problem across possibly weakly coupled subsystems, assigning a local controller to each subsystem. This leads to decentralised and distributed control architectures \cite{Siljak,SCATTOLINI2009723}.\\
Within the framework of model predictive control (MPC), numerous decentralised and distributed schemes have been proposed \cite{SCATTOLINI2009723} to replace a possibly large-scale centralised MPC with several smaller-scales ones. These methods can be broadly classified into cooperative and non-cooperative approaches. \\
In cooperative MPC \cite{STEWART2010460}, each local controller optimises a common plant-wide objective function. 
Under suitable assumptions, cooperative approaches can recover the solution to the centralised optimisation problem. However, optimality is achieved only asymptotically and requires multiple controllers' iterations and data exchange within each sampling instant. These requirements may significantly increase the computational and communication burden, potentially undermining the advantages of distributed architectures.\\
In contrast, non-cooperative MPC schemes allow each subsystem to compute its local control action independently, without requiring iterative coordination among controllers. Depending on the information shared among the controllers, these approaches can be either decentralised~\cite{Riverso_pnp,RIVERSO20142179} or distributed~\cite{Camponogara_distributed,Trodden01082010,FARINA20121088,farina2018distributed}. An inherent complexity of non-cooperative approaches lies in defining the design ingredients and deriving suitable robust positively invariant (RPI) sets~\cite{rakovic2005invariant}, commonly used to account for the coupling effects among subsystems. Consequently, existing robust non-cooperative MPC schemes have been primarily developed for linear models, limiting their applicability to plants characterised by complex nonlinear dynamics.

To extend the applicability of this approach to nonlinear models, this work focuses on recurrent neural network (RNN) models \cite{BONASSI202292}. RNNs are a class of neural networks that, thanks to their internal recurrent units, can be used as black-box identified models to capture broad classes of nonlinear dynamics. Besides their modelling capabilities, their structure can be exploited to replicate the modularity of the underlying plant \cite{deGiuli,ravasio2026physics} and derive tractable controller design conditions \cite{d2023incremental,ravasio2024lmi,ravasio2025recurrent}.\\
Motivated by these considerations, this work proposes a novel decentralised NMPC framework for networked systems described by input-coupled RNN models. The proposed approach exploits the well-established concept of incremental input-to-state stability ($\delta$ISS \cite{bayer2013discrete}) to design the local controllers and account for the effects of subsystem interconnections. In the proposed scheme, each subsystem computes its local control action independently by solving a local robust nonlinear MPC (NMPC) problem inspired by \cite{ravasio2024lmi}, while only sharing its local setpoint equilibrium condition with its neighbours. Furthermore, the local control ingredients are designed using a computationally tractable procedure based on linear matrix inequalities (LMIs \cite{boyd1994linear}).\smallskip\\
The paper is structured as follows. Section~\ref{sec:problem_statement} formulates the control problem. Sections~\ref{sec:DMPC} and~\ref{sec:DMPC_ingredients} present the decentralised NMPC algorithm and the design procedure, respectively. Simulation results and concluding remarks are presented in Sections \ref{sec:simulations} and \ref{sec:conclusion}, respectively. For clarity of exposition, the proof of the main results is reported in the Appendix.
\subsection*{Preliminaries}
The set of real numbers is denoted by $\mathbb{R}$, while $\mathbb{R}_{\geq 0}$ denotes the set of non-negative real numbers, and $\mathbb{R}_+ \coloneq \mathbb{R}_{\geq 0} \setminus \{0\}$ denotes the set of strictly positive real numbers. Given a vector $v \in \mathbb{R}^n$, $v^\top$ denotes its transpose, and $v^{(i)}$ its $i$-th entry. Given a matrix $M \in \mathbb{R}^{n \times n}$, its $i$-th row is denoted by $
M^{(i)}$, and its entry in position $(i,j)$ is denoted by $
m^{(i,j)}$. Given $n$ matrices $M_1,\dots,M_n$, we denote by $\diag(M_1,\dots,M_n)$ the block-diagonal matrix with $M_1,\dots,M_n$ on its main diagonal blocks. The matrix $I_n$ denotes the $n \times n$ identity matrix.
We define the sets $\mathbb{S}_+^n \coloneq \{ M \in \mathbb{R}^{n \times n} \,:\, M = M^\top \succ 0 \}$, $\mathbb{D}_+^n \coloneq \{ M\in\mathbb{S}_+^n  \,:\, m_{ij} = 0 \ \forall i \neq j \}$, and $\mathbb{D}_{\geq 0}^n \coloneq \{ M \in \mathbb{R}^{n \times n} \,:\, M \succeq 0 \text{ and } m_{ij} = 0 \ \forall i \neq j \}$.
Given a matrix $Q \in \mathbb{S}_+^n$, we define the ellipsoidal set $\mathcal{E}(Q)$ as $\mathcal{E}(Q) = \{ v \in \mathbb{R}^n \,:\, v^\top Q v \leq 1 \}$.
The sequence $\{u(k), \dots, u(k+N)\}$ is denoted compactly as $u(k:k+N)$. Given an index set $\mathcal{I} \subseteq \{1, \dots, n\}$, let $|\mathcal{I}|$ denote the cardinality of $\mathcal{I}$. Given two sets $\mathcal{A},\mathcal{B}\subseteq\mathbb{R}^n$, the Minkowski sum and Pontryagin set difference are defined as $\mathcal{A}\oplus\mathcal{B}:=\{a+b\,:\, a\in\mathcal{A},\,b\in\mathcal{B}\}$ and $\mathcal{A}\ominus\mathcal{B}:=\{x\in\mathbb{R}^n\,:\, x+\mathcal{B}\subseteq\mathcal{A}\}$, respectively. A continuous function $\alpha:\R_{\geq0}\to\R_{\geq0}$ is a class $\mathcal K$-function if $\alpha(s)>0$ for all $s>0$, it is strictly increasing and $\alpha(0)=0$. Also, a continuous function $\alpha:\R_{\geq0}\to\R_{\geq0}$ is a class $\mathcal K_\infty$-function if it is a class $\mathcal K$-function and $\alpha(s)\to\infty$ as $s\to\infty$. Finally, a continuous function $\beta:\R_{\geq0}\times\mathbb Z_{\geq0}\to\R_{\geq0}$ is a class $\mathcal{KL}$-function if $\beta(s,k)$ is a class $\mathcal K$-function with respect to $s$ for all $k$, it is strictly decreasing in $k$ for all $s>0$, and $\beta(s,k)\to0$ as $k\to\infty$ for all $s>0$. In the text, we make reference to the definition of globally Lipschitz continuous function, available in \cite{cao2004absolute}.
\smallskip\\
Consider a generic nonlinear discrete-time system described by
\begin{equation}\label{eq:nonlinear_sys}
    x(k+1)=f(x(k),w(k))\\
\end{equation}
where $k\in\mathbb Z_{\geq0}$ is the discrete-time index, $x\in\R^n$ is the state, and $w\in\R^m$ is a perturbation acting on the system.
We recall the following definitions.
\begin{definition}[RPI set~\cite{rakovic2005invariant}]
    A set $\Xi$ is an RPI set for system~\eqref{eq:nonlinear_sys}, if $x(k)\in\Xi$ implies $x(k+1)\in\Xi$, for all $w\in\mathcal W$.
\end{definition}
\begin{definition}[$\delta$ISS, \cite{bayer2013discrete}]\label{def:delta_ISS}
    System~\eqref{eq:nonlinear_sys} is said to be $\delta$ISS with respect to the sets $\mathcal{X}\subseteq\mathbb{R}^n$ and $\mathcal{W}\subseteq\mathbb{R}^m$ if $\mathcal{X}$ is RPI for \eqref{eq:nonlinear_sys} and there exist functions $\beta\in\mathcal{KL}$ and $\gamma\in\mathcal{K_\infty}$ such that for any $k\in\mathbb{Z}_{\geq0}$, any pair of initial states $x_a(0),x_b(0)\in\mathcal{X}$ and any pair of input $w_a\in\mathcal{W}$ and $w_b\in\mathcal{W}$, it holds that
    \begin{multline*}
    \norm{x_a(k)-x_b(k)}\leq\beta(\norm{x_a(0)-x_b(0)},k)\\+\gamma(\max_{h\geq 0}\norm{w_a(h)-w_b(h)}).
    \end{multline*}
\end{definition}
\vspace{0.5em}
In the following, since we consider global $\delta$ISS properties, i.e., with respect to $\mathbb{R}^n$ and $\mathcal W$, we simply refer to \eqref{eq:nonlinear_sys} as $\delta$ISS.
%
%
%
%
\section{PROBLEM STATEMENT}\label{sec:problem_statement}
\subsection{Plant decomposition}
Consider a plant composed of $M$ subsystems. 
We assume the dynamics of subsystem $i\in\mathcal M\coloneq\{1,\dots,M\}$ is described by the following RNN model
\begin{equation}\label{eq:subsystem_dynamics_0}
\begin{cases}
x_i(k{+}1)\!=\!f_i(A_ix_i(k)\!+B_{ii}u_i(k)\!+\!\sum_{j\in\mathcal N_i} B_{ij}u_j(k))\\
y_i(k)=C_ix_i(k)
\end{cases}\ 
\end{equation}
where $x_i\in\R^{n_i}$ represent the local state vector, $u_i\in\mathbb R^{m_i}$ the local input vector, and $y_i\in\mathbb R^{p_i}$ the local output vector. Matrices $A_i \in \mathbb{R}^{n_i \times n_i}$, $B_{ii} \in \mathbb{R}^{n_i \times m_i}$, $B_{ij} \in \mathbb{R}^{n_i \times m_j}$, and $C_{i} \in \mathbb{R}^{p_i \times n_i}$.\\
Model \eqref{eq:subsystem_dynamics_0} is denoted input-coupled in view of the particular interconnection with neighbouring subsystems $j\in\mathcal N_i\subseteq\mathcal M$, where, by definition, $i\notin\mathcal N_i$.\\
Notice that input-coupled subsystems arise naturally in relevant practical applications, including, e.g., water distribution systems \cite{cantoni2007control}. In addition, a centralised RNN model of the considered model class can always be decomposed into $M$ input-coupled subsystems of the form \eqref{eq:subsystem_dynamics} by applying the procedure presented in Appendix \ref{app:overlapping_decomposition}.\\
The function $f_i(\cdot)=\left[f_i^{(1)}(\cdot),\;\dots,\;f_i^{(n_i)}(\cdot)\right]^{\top}$, for $i=1,\dots,M$, is a vector of scalar functions applied element-wise satisfying the following assumption.
\begin{ass}\label{as:nonlinear_function}
For each $i\in\mathcal{M}$ and $r=1,\dots,n_i$, the function $f_i^{(r)}$ is either a nonlinear globally Lipschitz continuous function with Lipschitz constant $L_{\mathrm c,r}$ or the identity function.
\end{ass}
\vspace{0.5em}
Our goal is to develop, for each subsystem $i$, a local controller that computes the local control action $u_i$ to steer the output $y_i$ towards a desired setpoint $\bar y_i\in\R^{p_i}$, while accounting for the coupling effects induced by the inputs of the other subsystems. In addition, the control input and output of each subsystem need
to satisfy local constraints, i.e.,  $(u_i,y_i)\in\mathbb U_i\times\mathbb Y_i$.\\
We make the following assumption.
\begin{ass}
    The setpoints $\bar y_i\in\mathbb Y_i$, $i=1,\dots,M$, are defined such that there exist $M$ associated equilibrium tuples $(\bar u_i,\bar x_i)$ satisfying $\bar u_i\in\mathbb U_i$ and  $\bar x_i=f_i(A_i\bar x_i+B_{ii}\bar u_i+\sum_{j\in\mathcal N_i} B_{ij}\bar u_j)$.
\end{ass}
\vspace{0.5em}
Without loss of generality, we denote $\mathbb U_i\coloneq\{u_i\in\R^{m_i}\,:\,(u_i-\bar u_i)^\top Q_{u,i}^0(u_i-\bar u_i)\leq 1\}$, where $Q_{u,i}^0\in\mathbb D_+^{m_i}$, whereas the set $\mathbb Y_i\coloneq\{y_i\in\R^{p_i}\,:\,G_iy_i\leq b_i\}$ is polytopic, where $G_i\in\R^{n_{\mathrm r,i}\times p_i}$ and $b_i\in\R^{n_{\mathrm r,i}}$.\\
We assume that the input equilibria $\bar u_j$ are known to all subsystems $i$ such that $u_j\in\mathcal N_i$.
\subsection{The proposed strategy}
We reformulate the state dynamics of \eqref{eq:subsystem_dynamics_0} as
\begin{equation}\label{eq:subsystem_dynamics}
x_i(k{+}1){=}f_i(A_ix_i(k){+}B_{ii}u_i(k){+}\sum_{j\in\mathcal N_i}B_{ij}\bar u_j{+}w_i(k))
\end{equation}
where
\begin{equation}\label{eq:disturbance_bound}
w_i(k)\coloneq\sum_{j\in\mathcal N_i}B_{ij}(u_j(k)-\bar u_j),
\end{equation}
can be treated as an unknown disturbance capturing the coupling effects of the other subsystems on subsystem $i$. Since the input $u_j(k)\in\mathbb U_j$ for all $j\in \mathcal M$, it follows that $u_j(k)-\bar u_j\in\mathcal E(Q_{u,j}^0)$ for all $k\geq0$. Therefore, the disturbance $w_i(k)$ is also bounded, i.e., we can define the set $\mathbb W_i$ such that
\[w_i(k)\in\bigoplus_{j\in\mathcal N_i} B_{ij}\mathcal E(Q_{u,j}^0)\subseteq\mathbb W_i,\]
for all $k\geq 0$.\\ 
Inspired by \cite{riverso2013plug}, the key idea of the proposed approach is to control each subsystem using a local nonlinear model predictive controller, denoted by $i$-NMPC. 
Each $i$-NMPC is designed based on the algorithm presented in \cite{ravasio2024lmi} to ensure the robust convergence of subsystem $i$ to its local setpoint in the presence of the bounded disturbance $w_i(k)\in\mathbb W_i$.\\
In the following section, we present the algorithm under the state-feedback assumption to simplify the framework. When the system state is not available, an estimate of $x_i(k)$ for subsystem $i$ can be obtained using an observer. Provided that the control inputs can be shared among neighbouring subsystems, a local observer can be independently designed for each subsystem based on the approach in~\cite{d2023data}.
%
%
%
%
\section{The decentralised NMPC algorithm}\label{sec:DMPC}
\subsection{The local NMPC problem}
The definition of the $i$-NMPC problem requires the preliminary definition, for each subsystem $i$, of the associated nominal state dynamics
\begin{equation}\label{eq:subsystem_dynamics_nominal}
        \hat x_i(k+1)=f_i(A_i\hat x_i(k)\!+\!B_{ii}\hat u_i(k){+}\sum_{j\in\mathcal N_i}\!B_{ij}\bar u_j)
\end{equation}
where $\hat x_i\in\R^{n_i}$ and $\hat u_i\in\R^{m_i}$ represent the nominal state and input, respectively.\\
The control law for the $i$-th subsystem is defined as
\begin{equation}\label{eq:control_law_RPI}
    u_i(k)=\hat u_i(k)+K_i(x_i(k)-\hat x_i(k))
\end{equation}
where, as discussed below, $K_i\in\R^{m_i\times n_i}$ is the local control gain to be defined in order to impart closed-loop $\delta$ISS properties to \eqref{eq:subsystem_dynamics}.\\
Considering \eqref{eq:subsystem_dynamics} and \eqref{eq:control_law_RPI}, we obtain
\begin{multline}\label{eq:subsystem_dynamics_closed_loop}
        x_i(k+1)\!=\!f_i(A_{\mathrm K,i} x_i(k)\!+\!B_{ii}(\hat u_i(k)\!-\!K_i\hat x_i(k))\\+\!\sum_{j\in\mathcal N_i}\!B_{ij}\bar u_j\!+\!w_i(k))
\end{multline}
where $A_{\mathrm K,i}\coloneq A_i+B_{ii}K_i$.\\
Notice that the $\delta$ISS of \eqref{eq:subsystem_dynamics_closed_loop} enables the definition of an RPI set $\delta\mathbb X_i\subseteq\R^{n_i}$, such that if $\Delta x_i(k)\coloneq x_i(k)-\hat x_i(k)\in\delta\mathbb X_i$ and $w_i(k)\in\mathbb W_i$ then $\Delta x_i(k+1)\in\delta\mathbb X_i$.\\
As in standard tube-based NMPC, we define the tightened constraint sets $\hat{\mathbb Y}_i\coloneq\mathbb Y_i\ominus C_i\delta\mathbb X_i$ and $\hat{\mathbb U}_i\coloneq\mathbb U_i\ominus K_i\delta\mathbb X_i$. We make the following assumption.
\begin{ass}\label{ass:constr_tightening}
The sets $\hat{\mathbb Y}_i$ and $\hat{\mathbb U}_i$ are nonempty, and
$\bar y_i\in\hat{\mathbb Y}_i$ and $\bar u_i\in\hat{\mathbb U}_i$.
\end{ass}
\vspace{0.5em}
The nominal $i$-NMPC problem is formulated as follows.
\begin{subequations}\label{opt:i-MPC}
    \begin{align}
        \mathbb P_i(x(k))\coloneq&\min_{\hat x_i(0),\hat u_i(0:N-1)}J_i(\hat x_i(0),\hat u_i(0:N-1))\\
        &\text{subject to}\notag\\
        &x_i(k)-\hat x_i(0) \in \delta\mathbb{X}_i,\label{opt:init}\\
&\forall\tau=0,\dots,\ N-1:\notag\\
&\ \  \ \hat{x}_i(\tau{+}1)\!=\!f_i(A_i\hat{x}_i(\tau)\!+\!B_{ii}\hat{u}_i(\tau)\!+\!\sum_{j\in\mathcal N_i}\!B_{ij}\bar u_j),\notag\\
&\ \ \ \hat{u}_i(\tau)\in \hat{\mathbb U}_i,\; C_i\hat{x}_i(\tau)\in\hat{\mathbb Y}_i,\notag \\
&\hat{x}_i(N) \in \mathbb X_{\mathrm f,i}\label{opt:terminal_set}      
\end{align}
\end{subequations}
where $N\in\mathbb Z_+$ is the prediction horizon, $\mathbb X_{\mathrm f,i}\subseteq\R^{n_i}$ is the terminal set that will be specified below. The cost function $J_i$ is defined as
\[J_i{\coloneq}\sum_{\tau=0}^{N{-}1}l_i(\hat x_i(\tau),\hat u_i(\tau))\!+\!V_{\mathrm f,i}(\hat x_i(N)),\]
where $V_{\mathrm f,i}:\R^{n_i}\to\R_{\geq0}$ is the terminal cost, and the stage cost
\(l_i(\hat x_i(k),\hat u_i(k))\coloneq\norm{\hat x_i(k){-}\bar x}_{\Lambda_{\mathrm x,i}}^2+\norm{\hat u_i(k)-\bar u_i}_{\Lambda_{\mathrm u,i}}^2,\) where $\Lambda_{\mathrm x,i}\in\mathbb D_+^{n_i}$ and $\Lambda_{\mathrm u,i}\in\mathbb D_+^{m_i}$.\\
Let the solution to the $i$-NMPC \eqref{opt:i-MPC} at time $k$ be $(\hat{x}_i(0|k),\hat{u}_i(0:N-1|k))$. According to \eqref{eq:control_law_RPI}, the control input applied to subsystem $i$ is  $u_i(k)=\hat{u}_i(0|k)+K_i(x_i(k)-\hat{x}_i(0|k))$.
\subsection{Terminal ingredients}
Regarding the choice of the nominal $i$-NMPC terminal ingredients, we introduce, as in \cite{ravasio2024lmi}, the auxiliary control law
\begin{equation}\label{eq:auxiliary_law}
\hat{u}_i(k)=K_i^\mathrm{aux}(\hat x_i(k)-\bar x_i)+\bar u_i.
\end{equation}
where the control gain $K_i^\mathrm{aux}\in\R^{m_i\times n_i}$.\\
%
We select the terminal cost as
\[V_{\mathrm f,i}(\hat x_i(k))=\norm{\hat x_i(k)-\bar x_i}_{S_i}^2,\]  
and the terminal set as
\begin{equation*}
\mathbb X_{\mathrm f,i} =\{\hat{x}_i\in \R^{n_i}:(\hat{x}_i-\bar x_i)^\top S_i (\hat{x}_i-\bar x_i)\leq\gamma_{\mathrm f,i}\},
\end{equation*}
where $S_i\in\mathbb D_+^{n_i}$ and $\gamma_{\mathrm f,i}\in\R_+$ are design parameters to be selected in order to satisfy the following assumption.

\begin{ass}\label{ass:iNMPC_conditions}
For all $i\in\mathcal M$, the following conditions hold
\begin{itemize}

\item[\emph{(i)}]
for all $\hat x_i\in\mathbb X_{\mathrm f,i}$, $\bar u_i{+}K_i^{\mathrm{aux}}(\hat x_i-\bar x_i){\in}\hat{\mathbb U}_i$ and
$C_i\hat x_i{\in}\hat{\mathbb Y}_i$;


\item[\emph{(ii)}] matrix $S_i\in\mathbb D_+^{n_i}$ is structured such that its off-diagonal entries
$s_i^{(l,m)}$ are zero for all $l$ such that
$f_i^{(l)}(\cdot)\neq\mathrm{id}(\cdot)$ and $m\neq l$, and satisfies
\begin{multline}\label{eq:lyapunov_inequality_design}
S_i{-}(\tilde A_i{+}\tilde B_{ii}K_i^\mathrm{aux})^\top S_i(\tilde A_i{+}\tilde B_{ii}K_i^\mathrm{aux})\succeq\\\Lambda_{\mathrm x,i}{+}{K_i^\mathrm{aux}}^\top\Lambda_{\mathrm u,i}K_i^\mathrm{aux}.
\end{multline}
\end{itemize}
\end{ass}
\vspace{0.5em}
\subsection{Main result}
Before stating the main result, we define the collective variables $\mathbf x(k)\coloneq[x_1^\top(k),\;\dots,\;x_M^\top(k)]^\top$, $\hat{\mathbf x}(k)\coloneq[\hat x_1^\top(k),\;\dots,\;\hat x_M^\top(k)]^\top$, $\hat{\mathbf u}(k)\coloneq[\hat u_1^\top(k),\;\dots,\;\hat u_M^\top(k)]^\top$, $\mathbf y(k)\coloneq[y_1^\top(k),\;\dots,\;y_M^\top(k)]^\top$, $\bar{\mathbf{x}}\coloneq[\bar x_1^\top,\;\dots,\;\bar x_M^\top]^\top$, $\bar{\mathbf{u}}\coloneq[\bar u_1^\top,\;\dots,\;\bar u_M^\top]^\top$, and $\bar{\mathbf{y}}\coloneq[\bar y_1^\top,\;\dots,\;\bar y_M^\top]^\top$, function $f(\cdot)=[f_1(\cdot)^\top,\;\dots,\;f_M(\cdot)^\top]^\top$, and matrices $\mathbf K\coloneq\diag(K_1,\dots,K_M)$, $\mathbf A\coloneq\diag(A_1,\dots,A_M)$, $\mathbf C\coloneq\diag(C_1,\dots,C_M)$, and 
\[\mathbf B\coloneq\begin{bmatrix}
    B_{11}&\dots&B_{1M}\\
    \vdots&\ddots&\vdots\\
    B_{M1}&\dots&B_{MM}
\end{bmatrix}.\]
The evolution of the collective system given by \eqref{eq:subsystem_dynamics} and \eqref{eq:control_law_RPI} is
\begin{equation}\label{eq:full_system_dynamics}
    \begin{cases}
        \mathbf x(k{+}1){=}f((\mathbf A+\mathbf B\mathbf K)\mathbf x(k){+}\mathbf B(\hat{\mathbf u}(k){-}\mathbf K\hat{\mathbf x}(k))\\
        \mathbf y(k)=\mathbf C\mathbf x(k)
    \end{cases}
\end{equation}
We define the control gains in such a way that the following assumption is verified.
\begin{ass}\label{ass:RPI_conditions}
    The local control gains $K_1$, $\dots$, $K_M$ are defined such that
    \begin{itemize}
        \item[\emph{(i)}] for all $i\in\mathcal M$, the local closed-loop dynamics \eqref{eq:subsystem_dynamics_closed_loop} is $\delta$ISS, and $\delta\mathbb X_i$ is a compact RPI set for $\Delta x_i(k)$;
        \item[\emph{(ii)}] the collective closed-loop dynamics \eqref{eq:full_system_dynamics} is $\delta$ISS.
    \end{itemize}
\end{ass}
\vspace{0.5em}
The following theorem presents the main result.
\begin{theorem}\label{th:main_result}
    Let Assumptions \ref{as:nonlinear_function}-\ref{ass:RPI_conditions} hold. Define the feasibility region of the $i$-NMPC problem as
    \[\mathbb X_i^{\mathrm{NMPC}}\coloneq\{z_i\in \R^{n_i}\,:\,\eqref{opt:i-MPC}\text{ is feasible for } x_i(k)=z_i\},\]
    and the collective feasibility region as $\mathbb X^{\mathrm {NMPC}}\coloneq\prod_{i\in\mathcal M}\mathbb X_i^{\mathrm{NMPC}}$.\\
    Then,  
\begin{enumerate}
    \item[\emph{(i)}] if $\mathbf x(0)\in\mathbb X^{\mathrm {NMPC}}$, i.e., if $x_i(0)\in\mathbb X_i^{\mathrm{NMPC}}$ for all
$i\in\mathcal M$, the nominal $i$-NMPC problem \eqref{opt:i-MPC} remains
    feasible for all $i\in\mathcal M$ and $k\geq0$;
    \item[\emph{(ii)}] the equilibrium
    $(\bar{\mathbf x},\bar{\mathbf u},\bar{\mathbf y})$
    is asymptotically stable for the closed-loop system \eqref{eq:full_system_dynamics} and $\mathbb X_i^{\mathrm{NMPC}}$ is a region of attraction.
    \end{enumerate}
\end{theorem}
\vspace{0.5em}
The proof of Theorem~\ref{th:main_result} is reported in Appendix~\ref{app:proofs}.
%
%
%
%
%
\section{Synthesis of the decentralised NMPC}\label{sec:DMPC_ingredients} 
The design of the proposed decentralised NMPC scheme requires the definition, for each $i\in\mathcal M$, of $(K_i^\mathrm{aux},S_i,\gamma_{\mathrm f,i})$ satisfying Assumption~\ref{ass:iNMPC_conditions} and $(K_i,P_i,\gamma_i)$ satisfying Assumption~\ref{ass:RPI_conditions}.\\
This section provides design conditions for the definition of these
ingredients.
\subsection{Local control gain and RPI set}
The following proposition provides sufficient conditions for computing
the local control gains $K_i$ and the corresponding RPI sets
$\delta\mathbb X_i$ satisfying Assumption~\ref{ass:RPI_conditions}-\emph{(i)}. Before stating it, define the matrices $\tilde A_i\coloneq W_iA_i$ and $\tilde B_{ii}\coloneq W_iB_{ii}$, where $W_i\in\mathbb S_+^{n_i}$ is defined such that $w_i^{(r,r)}=L_{\mathrm c,r}$ if $f_i^{(r)}\neq\mathrm{id}(\cdot)$ and $w_i^{(r,r)}=1$ otherwise. We also denote the number of neighbours of subsystem $i$ by $\nu_i\coloneq|\mathcal N_i|$.
\begin{proposition}\label{prop:diss_desing}
    Consider the local closed-loop dynamics~\eqref{eq:subsystem_dynamics_closed_loop}, and let Assumption~\ref{as:nonlinear_function} hold. If there exist a matrix $Q_i\in\mathbb S_+^{n_i}$ structured in such a way that its off-diagonal entries $q_i^{(r,s)}$ are zero for all $r$ where $f_i^{(r)}(\cdot) \neq \mathrm{id}(\cdot)$ and $r \neq s$, matrix $L_i\in\R^{n_i\times m_i}$, and scalars $\gamma_i,\tau_i\in\R_+$ such that
    \begin{equation}
    \begin{bmatrix}
        \left(1-\frac{1}{\gamma_i}\right)Q_i
    &
    Q_i\tilde A_i^{\top}
    +
    L_i^{\top}\tilde B_{ii}^{\top}
    \\
    \tilde A_i Q_i+\tilde B_{ii} L_i
    &
    \frac{1}{1+\tau_i^2}Q_i
    \end{bmatrix}\succeq 0\label{eq:RPI_cstr_1}
    \end{equation}
    then, setting $P_i=Q_i^{-1}$ and $K_i=L_iQ_i^{-1}$, the local closed-loop dynamics \eqref{eq:subsystem_dynamics_closed_loop} is $\delta$ISS with respect to $\tilde w_i(k)\coloneq B_{ii}(\hat u_i(k)-K_i\hat x_i(k))+\sum_{j\in\mathcal N_i}B_{ij}\bar u_j+w_i(k)$.\\ 
    Moreover, let $w_i(k)$ be given by~ \eqref{eq:disturbance_bound},  $u_j(k)\in\mathbb U_j$ for all $j\in\mathcal N_i$ and $k\geq0$, and the condition 
    \begin{equation}
    \begin{bmatrix}
        Q_{u,j}^0&\nu_iB_{ij}^\top\\
        \nu_iB_{ij}&\frac{\tau_i^2}{1+\tau_i^2}Q_i
    \end{bmatrix}\succeq0, \quad \forall j\in\mathcal N_i\label{eq:RPI_cstr_2}
    \end{equation}
Then, the set $\delta\mathbb X_i\coloneq\mathcal E(P_i/\gamma_i)$ is RPI for $\Delta x_i(k)$.
\end{proposition}
\vspace{0.5em}
The proof of Proposition~\ref{prop:diss_desing} is reported in Appendix~\ref{app:proofs}.\\
Notice that, the construction of the RPI set $\delta\mathbb X_i$ relies on the assumption that $u_j\in\mathbb U_j$ for all $j\in\mathcal N_i$, which is guaranteed by the corresponding $j$-NMPC problems. Moreover, once $K_i$ and $\delta\mathbb X_i$ have been computed, it must be checked whether Assumption~\ref{ass:constr_tightening} is satisfied by verifying that
$\mathbb U_i\ominus K_i\delta\mathbb X_i\neq\emptyset$
and
$\mathbb Y_i\ominus C_i\delta\mathbb X_i\neq\emptyset$ for all $i\in\mathcal M$.
\subsection{Offline design procedure}
Based on the above discussion, the procedure for selecting the ingredients of the proposed decentralised NMPC scheme is summarised in the following three steps.\smallskip\\
\emph{Step 1.}  For all $i\in\mathcal M$, initialise $\gamma_i>1$ and select a sufficiently large $\tau_i\in\mathbb R_+$. Then, solve the LMI problem
\begin{align*}
&\min_{\alpha_i,Q_i,L_i} \alpha_i\\
&\text{subject to}\\
&\quad\eqref{eq:RPI_cstr_1}-\eqref{eq:RPI_cstr_2},\\
&\quad f_\mathrm{LMI}(Q_i/\gamma_i)\preceq\alpha_i I,
\end{align*}
where $\alpha_i\in\mathbb R_+$ and $f_\mathrm{LMI}(\cdot)$ is a designer-selected function aimed at promoting a small RPI set (see \cite{boyd1994linear} for possible choices). If the problem is infeasible, increase $\gamma_i$ and/or reduce $\tau_i$, and solve the problem again. Once a feasible solution is obtained, set $K_i=L_iQ_i^{-1}$ and $P_i=Q_i^{-1}$. Finally, verify that Assumption~\ref{ass:constr_tightening} is satisfied.\\
\emph{Step 2.} Verify that there exists a matrix $\mathbf P\in\mathbb S_+^n$, where $n=\sum_{i\in\mathcal M}n_i$, structured such that its off-diagonal entries $\mathbf p^{(r,s)}$ are zero for all $r$ such that $f^{(r)}(\cdot)\neq\mathrm{id}(\cdot)$ and $r\neq s$, satisfying
\[\mathbf P-(\mathbf{WA}+\mathbf{WBK})^\top \mathbf P(\mathbf{WA}+\mathbf{WBK})\succ0,\] 
where $\mathbf W\coloneq\diag(W_1,\dots,W_M)$. This LMI condition guarantees
that the collective closed-loop dynamics \eqref{eq:full_system_dynamics} satisfies
Assumption~\ref{ass:RPI_conditions}-\emph{(ii)} (see \cite[Theorem~2]{d2023incremental}).\\
\emph{Step 3.} For all $i\in\mathcal M$, set $K_i^\mathrm{aux}=K_i$, fix the design parameters $\Lambda_{\mathrm x,i}\in\mathbb D_+^{n_i}$, $\Lambda_{\mathrm u,i}\in\mathbb D_+^{m_i}$ and $\varepsilon_i,\bar{\varepsilon}_i\in\mathbb R_+$, and solve the LMI problem
\begin{subequations}
\begin{align}
&\min_{S_i,\tilde{\gamma}_{\mathrm f,i}}\quad
\varepsilon_i \operatorname{trace}(S_i)
+ \bar{\varepsilon}_i \tilde{\gamma}_{\mathrm f,i} \label{opt:terminal_incredients_cost}\\
&\text{subject to}\nonumber\\
&\quad\eqref{eq:lyapunov_inequality_design},\nonumber\\
&\quad\begin{bmatrix}
S_i & (G_i^{(r)}C_i)^\top \\
G_i^{(r)}C_i &
\tilde{\gamma}_{\mathrm f,i}
\bigl(\hat b_i^{(r)} - G_i^{(r)}C_i\bar x_i\bigr)^2
\end{bmatrix}
\succeq 0,
\quad r=1,\ldots,n_{\mathrm r,i}\label{opt:cstr_output}\\
&\quad \tilde{\gamma}_{\mathrm f,i}\frac{ K_i^{\prime} Q_u^0K_i}
{\bar b_{i}}
\preceq S_i\label{opt:cstr_input}
\end{align}
\end{subequations}
where $\tilde \gamma_{\mathrm f,i}\in\mathbb R_+$, $\hat b_i^{(r)}\coloneq b_i^{(r)}-\max_{\Delta x_i\in\delta\mathbb X_i} |G_i^{(r)}C_i\Delta x_i|$, and $\bar b_{i}
=
2\left(
1-\frac{1}{2}
\max_{\Delta x_i\in\delta\mathbb{X}_i}
\left\|
\Delta x_i
\right\|_{K_i^\top Q_u^0 K_i}
\right)$. Then, set
$\gamma_{\mathrm f,i}=1/\tilde{\gamma}_{\mathrm f,i}$.\\
Notice that the cost function \eqref{opt:terminal_incredients_cost} is selected to
promote a large terminal set (see~\cite{boyd1994linear}), whereas constraints~\eqref{opt:cstr_output} and \eqref{opt:cstr_input} respectively guarantee that
$C_i\hat x_i(k)\in\hat{\mathbb Y}_i$ and $(\bar u_i{+}K_i(\hat x_i{-}\bar x_i))\in\hat{\mathbb U}_i$ for all
$(\hat x_i(k){-}\bar x_i)\in
\mathcal E(S_i/\gamma_{\mathrm f,i})$.
\section{Case study}\label{sec:simulations}
In this section, we apply the proposed decentralised scheme to the control of a simulated quadruple tank system.\\
The process, described in detail in \cite{johansson2000}, consists of four interconnected tanks and two controllable pumps, which supply water to the upper tanks with flow rates $q_1$ and $q_2$, respectively. The control objective is to regulate the water levels $h_1$ and $h_2$ in the two lower tanks. In the considered configuration, only these two levels are measured.\\
The control inputs and measured outputs are collected into the vectors
$u=[q_1,\;q_2]^\top$ and $y=[h_1,\;h_2]^\top$, respectively. The inputs and outputs are subject to the following constraints: \(q_1\in[0,\,6.42\cdot10^{-4}]~\mathrm{m^3/s}\), \(q_2\in[0,\,7.9\cdot10^{-4}]~\mathrm{m^3/s}\), and \(h_i\in[0,\,1.3]~\mathrm{m}\) for $i=1,2$.\\
To evaluate the proposed strategy, a first-principles model of the quadruple-tank system, as described in~\cite{johansson2000}, has been implemented in MATLAB.
\subsection{Modelling}
To obtain an RNN-based model suitable for distributed control, the plant simulator has been excited using multilevel pseudo-random input signals designed to explore different operating conditions and frequencies. Three independent input-output sequences of length $9000$, $3000$ and $1000$ have been obtained with a sampling time of $25~\mathrm s$ and used as training, validation, and test datasets, respectively.  White noise has been added to account for measurement disturbances. The resulting dataset has then been normalised before being used for model identification.\\
Following the echo-state learning procedure presented in~\cite{jaeger2002adaptive}, a centralised RNN model of the form~\eqref{eq:plant_dynamics_f}, with $n=16$ states, has been identified from the collected dataset. The activation functions are selected as $f^{(i)}=\tanh(\cdot)$ for $i=1,2,3$ and $f^{(i)}=\mathrm(\cdot)$ for $i=4,\dots,16$. The FIT index~\cite{BONASSI202292}, computed on the test dataset, is $90.1\%$, indicating good modelling accuracy.\\
The identified centralised model has been subsequently decomposed into two input-coupled submodels using the overlapping decomposition described in Appendix~\ref{app:overlapping_decomposition}.  For each submodel $i=1,2$,  the local control input is $u_i=q_i$ and the local output is $y_i=h_i$.
\subsection{Control}
\begin{figure}[htbp]
    \centering
    \begin{subfigure}{0.85\linewidth}
        \centering
        \includegraphics[width=\linewidth]{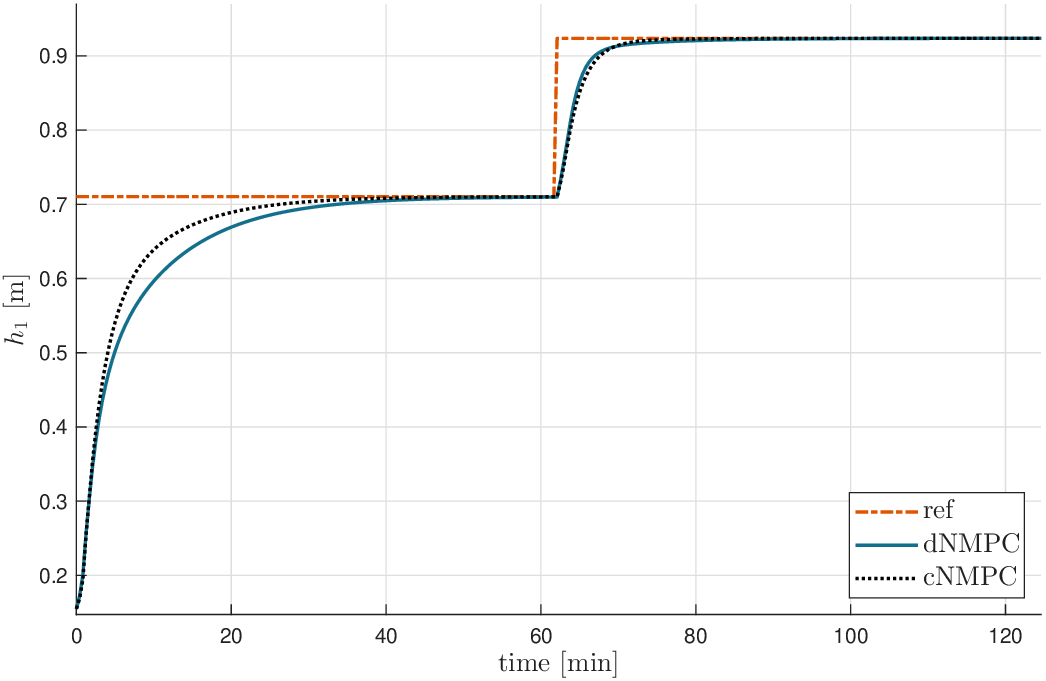}
    \end{subfigure}
    \hfill
    \begin{subfigure}{0.85\linewidth}
        \centering
        \includegraphics[width=\linewidth]{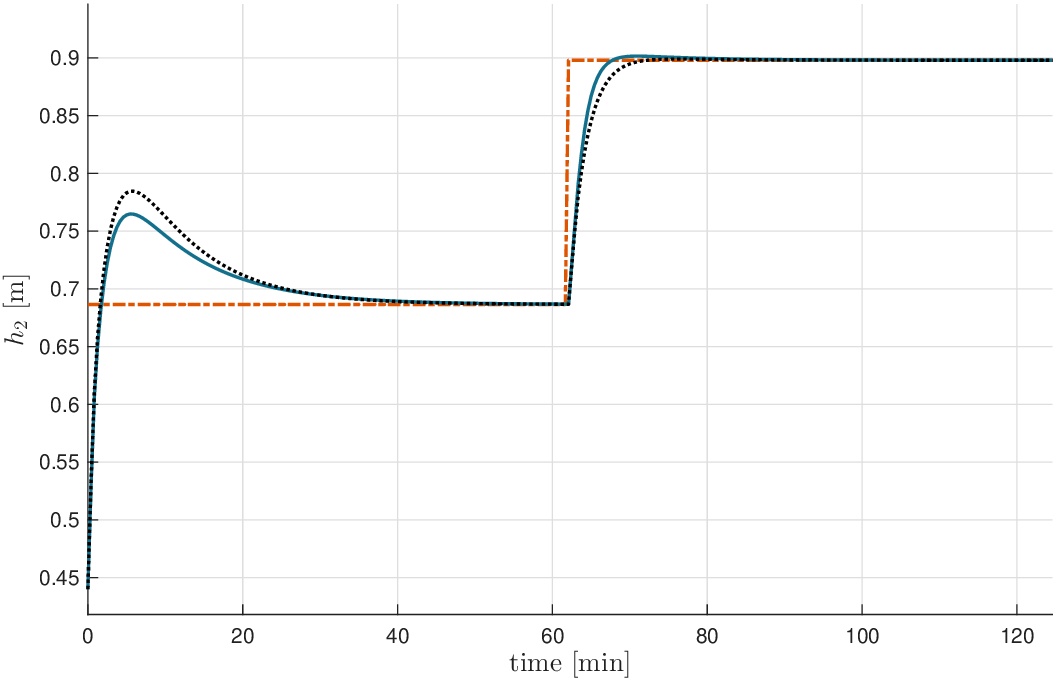}
    \end{subfigure}
    \caption{Closed-loop tracking performance: comparison between the centralised and decentralised NMPC schemes.}
    \label{fig:outputs_evolution}
\end{figure}

\begin{figure}[htbp]
    \centering
    \begin{subfigure}{0.85\linewidth}
        \centering
        \includegraphics[width=\linewidth]{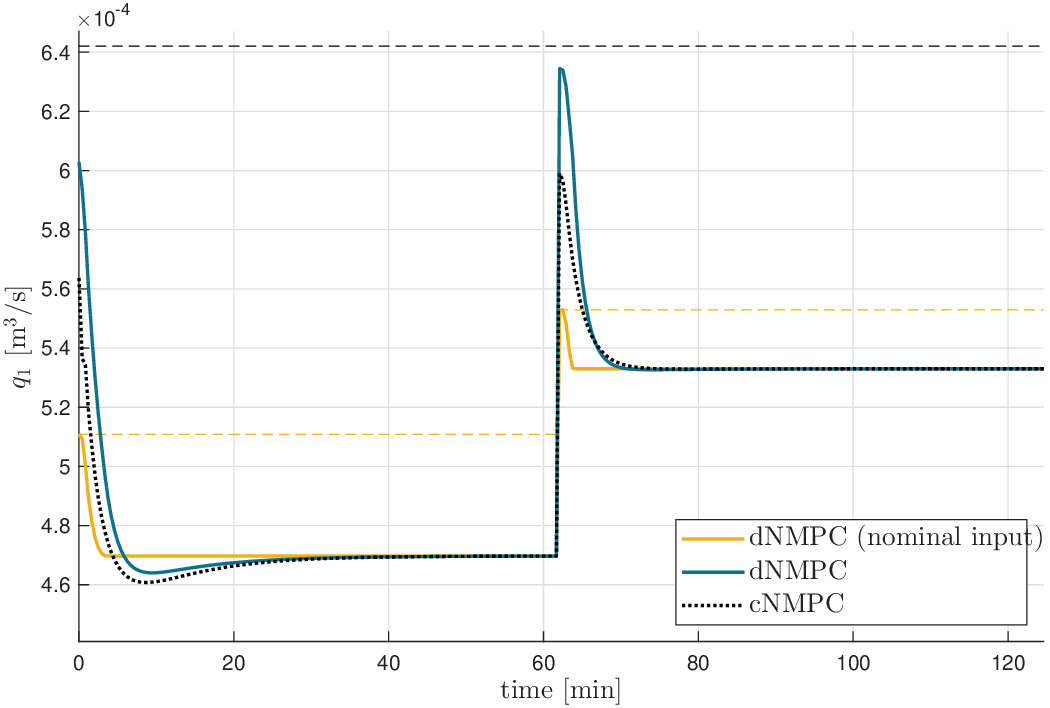}
    \end{subfigure}
    \hfill
    \begin{subfigure}{0.85\linewidth}
        \centering
        \includegraphics[width=\linewidth]{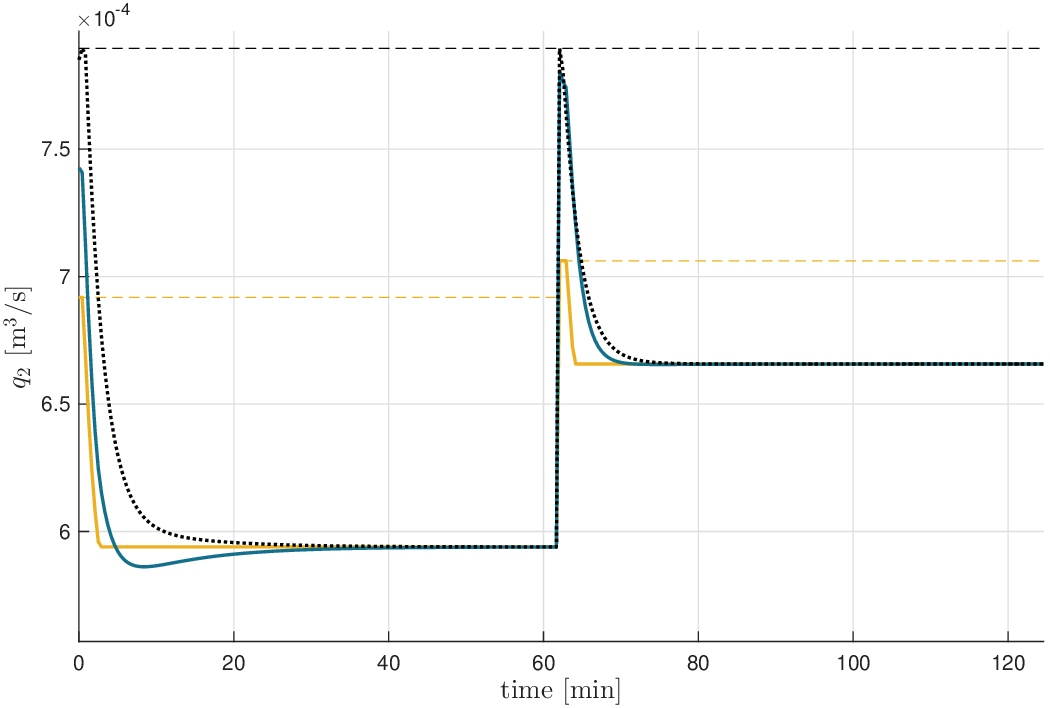}
    \end{subfigure}
    \caption{Control input evolution: comparison between the centralised and decentralised NMPC schemes. The black dashed lines represent the constraints on $u$, while the yellow dashed lines represent the constraints on $\hat{u}$.}
    \label{fig:input_evolution}
\end{figure}

Using the identified input-coupled RNN model, the proposed decentralised control scheme has been implemented and evaluated in simulation.\\
The $i$-NMPC weight matrices have been selected as $\Lambda_{x,i}=10\cdot I_{n_i}$ and $\Lambda_{u,i}=I_{m_i}$, for $i=1,2$, and the prediction horizon as $N=25$.\\
To assess the effectiveness of the proposed approach, the performance of the decentralised controller (dNMPC) has been compared with that of a centralised nominal nonlinear model predictive controller (cNMPC), designed using the identified centralised RNN-based model. The simulation results are reported in Figures~\ref{fig:outputs_evolution}-\ref{fig:input_evolution}.\\
As shown in the figures, the setpoints are successfully tracked, while input and output constraints are satisfied throughout the simulation. Overall, the proposed decentralised scheme achieves closed-loop performance comparable to that of the centralised controller, although some performance degradation arises from the conservative nature of the robust design. Figure~\ref{fig:input_evolution} shows that, when $\hat u_1(k)$ and $\hat u_2(k)$ (yellow lines) reach their respective tightened constraint boundaries, the real control inputs generated by the dNMPC according to \eqref{eq:control_law_RPI} do not reach the boundaries of the original feasibility sets. This is a consequence of the constraint tightening based on RPI sets. To quantify the resulting loss of optimality, we have computed the following cumulative cost over the entire discrete-time simulation horizon $T_\mathrm s=300$ steps:
\[
J_{\mathrm{s}}=\sum_{t=0}^{T_{\mathrm s}-1}\left(
\norm{x(t)-\bar{x}(t)}_{\Lambda_x}^2
+
\norm{u(t)-\bar{u}(t)}^2_{\Lambda_u}
\right),
\]
where $\Lambda_x=\operatorname{diag}(\Lambda_{x,1},\Lambda_{x,2})$ and $\Lambda_u=\operatorname{diag}(\Lambda_{u,1},\Lambda_{u,2})$. The resulting values are $J_{\mathrm{s}}=104.9$ for the cNMPC and $J_{\mathrm{s}}=113.2$ for the dNMPC, corresponding to a performance loss of $7.9\%$ with respect to the cNMPC.
%
%
%
%
%
\section{Conclusions}\label{sec:conclusion}
In this work, we have presented a decentralised algorithm for systems described by input-coupled RNN models, extending the approach in \cite{RIVERSO20142179} to nonlinear models. Future work will focus on extending the proposed approach to more general RNN models, possibly including state couplings, and on developing methods to distribute the computation of the control ingredients among the subsystems.
%
%
\appendix
\subsection{Proof of the main results}\label{app:proofs}
\textit{Proof of Theorem \ref{th:main_result}.}
The proof of Theorem \ref{th:main_result} follows similar lines as the proof of \cite[Theorem~1]{riverso2013plug}.\\
    \emph{(i)} Consider the closed-loop nominal dynamics given by~\eqref{eq:subsystem_dynamics_nominal} and~\eqref{eq:auxiliary_law}, i.e., 
    \begin{equation}\label{eq:nominal_closed_loop_subsystem_dynamics}
        \hat x_i(k+1)=f_i((A_i+B_{ii}K_i^\mathrm{aux})\hat x_i(k)+\bar w_i)
    \end{equation}
    where $\bar w_i=B_{ii}(\bar u_i-K_i^\mathrm{aux}\bar x_i)$.\\
    Notice that, in view of Assumption~\ref{ass:iNMPC_conditions}-\emph{(ii)} and the fact that $\Lambda_{\mathrm x,i}{+}{K_i^\mathrm{aux}}^\top\Lambda_{\mathrm u,i}K_i^\mathrm{aux}\succ0$, it holds that 
    \begin{equation*}
    S_i{-}(\tilde A_i{+}\tilde B_{ii}K_i^\mathrm{aux})^\top S_i(\tilde A_i{+}\tilde B_{ii}K_i^\mathrm{aux})\succ0.
    \end{equation*}
    Using similar arguments as in the proof of~\cite[Theorem~1]{ravasio2024lmi}, this condition implies that the pair $(\bar x_i,\bar u_i)$ is an asymptotically stable equilibrium for \eqref{eq:nominal_closed_loop_subsystem_dynamics} with Lyapunov function $\hat V_{\mathrm f,i}(\hat x_i(k))=\norm{\hat x_i(k)-\bar x_i}^2_{S_i}$. In view of it, the set $\mathbb X_{\mathrm f,i}$ is positively invariant for \eqref{eq:nominal_closed_loop_subsystem_dynamics}.\\
    Now, define $\hat u_i^{\mathrm{aux}}(N|k)\coloneq\bar u_i+K_i^{\mathrm{aux}}(\hat x_i(N|k)-\bar x_i)$ and $\hat x_i^{\mathrm{aux}}(N+1|k)\coloneq\bar f_i((A_i+B_{ii}K_i^\mathrm{aux})\hat x_i(N|k)+\bar w_i)$. Also, define the sequences \begin{align*}
&\tilde u_i(1:N|k)\coloneq\{\hat u_i(1|k),\dots,\hat u_i(N-1|k),\hat u^{\mathrm{aux}}_i(N|k)\}\\
&\tilde x_i(1:N+1|k)\coloneq\{\hat x_i(1|k),\dots,\hat x_i(N|k),\hat x^{\mathrm{aux}}_i(N+1|k)\}.
\end{align*}
Recalling that $\delta\mathbb X_i$ is an RPI set for $\Delta x_i(k)$ and that $w_i(k)$ satisfies \eqref{eq:disturbance_bound}, constraint \eqref{opt:init} implies $x_i(k+1)-\tilde x_i(1|k)\in\delta\mathbb X_i$. 
Using standard arguments, it follows that $(\tilde x_i(1|k),\tilde u_i(1:N|k))$ provide a feasible solution for \eqref{opt:i-MPC} at time $k+1$.
\smallskip\\
\emph{(ii)} First, we prove convergence of the optimal cost function of the nominal $i$-NMPC problem.\\  
By optimality and in view of \emph{(i)}, it follows that
\begin{multline*}
\mathbb P_i(x(k+1))
\leq\sum_{\tau=1}^{N}l_i(\tilde x_i(\tau|k),\tilde 
u_i(\tau|k))+V_{\mathrm f,i}(\tilde x_i(N+1|k)).
\end{multline*}
Consider the closed-loop nominal dynamics \eqref{eq:nominal_closed_loop_subsystem_dynamics}. Using the same arguments as in the proof of~\cite[Theorem~1]{ravasio2024lmi}, Assumption \ref{ass:iNMPC_conditions}-\emph{(ii)} implies that
\begin{equation}\label{eq:lyapunov_decrease_iNMPC}
    V_{\mathrm f,i}(\hat x(k{+}1)){-}V_{\mathrm f,i}(\hat x(k)){\leq}{-}l(\hat x_i(k),K_i^{aux}(\hat x_i(k){-}\bar x_i){+}\bar u_i).
\end{equation}
Using \eqref{eq:lyapunov_decrease_iNMPC}, it holds that
\(V_{\mathrm f,i}(\tilde x_i(N{+}1|k)){-}V_{\mathrm f,i}(\tilde x_i(N|k))\leq {-}l_i(\tilde x_i(N|k),\tilde 
u_i(N|k)).\)
Therefore, it follows that
\begin{align*}
&\mathbb P_i(x(k+1))\leq\sum_{\tau=1}^{N-1}l_i(\tilde x_i(\tau|k),\tilde 
u_i(\tau|k))+V_{\mathrm f,i}(\tilde x_i(N|k))\\
&\quad=J_i(\hat x(0|k),\hat u_i(0:N-1|k))-l_i(\tilde x_i(0|k),\tilde 
u_i(0|k))
\end{align*}
Hence, the optimal cost satisfies
\[\mathbb P_i(x(k+1))-\mathbb P_i(x(k))\leq -l_i(\tilde x_i(0|k),\tilde 
u_i(0|k)),\]
and therefore $(\hat x_i(k),\hat u_i(k))\to(\bar x_i,\bar u_i)$ as $k\to+\infty$.\smallskip\\
Now, we consider the collective closed-loop dynamics \eqref{eq:full_system_dynamics}.\\
Define $\Delta\hat{\mathbf u}(k)\coloneq\hat{\mathbf u}(k)-\bar{\mathbf u}$, $\Delta\hat{\mathbf x}(k)\coloneq\hat{\mathbf x}(k)-\bar{\mathbf x}$, and $\hat{\mathbf w}(k)\coloneq \mathbf B(\Delta\hat{\mathbf u}(k)-\mathbf K\Delta\hat{\mathbf x}(k))$. Then, \eqref{eq:full_system_dynamics} can be rewritten as
\begin{equation}\label{eq:full_system_dynamics_2}
    \mathbf x(k+1)=f((\mathbf A+\mathbf{BK})\mathbf x(k)+\mathbf B(\mathbf{\bar u}-\mathbf{K}\bar {\mathbf x})+\hat{\mathbf w}(k)).
\end{equation}
By Assumption \ref{ass:RPI_conditions}-\emph{(ii)} and Definition \ref{def:delta_ISS}, there exist functions $\beta\in\mathcal{KL}$ and $\gamma\in\mathcal{K_\infty}$ such that, by comparing the actual closed-loop trajectory with the equilibrium trajectory, for any $k_0\geq0$ and $k\geq k_0$,
\begin{equation}\label{eq:diss_equilibrium}
\norm{\mathbf x(k)-\bar{\mathbf x}}
\leq
\beta(\norm{\mathbf x(k_0)-\bar{\mathbf x}},k-k_0)
+
\gamma(\max_{h\geq k_0}\norm{\hat{\mathbf w}(h)}).
\end{equation}
If $\mathbf x(0)\in\mathbb X^{\mathrm{NMPC}}$, the asymptotic convergence of $\hat x_i(k)$ and $\hat u_i(k)$ towards $\bar x_i$ and $\bar u_i$, respectively, for all $i\in\mathcal M$, implies $\hat{\mathbf x}(k)\to\bar{\mathbf x}$ and $\hat{\mathbf u}(k)\to\bar{\mathbf u}$, and hence $\hat{\mathbf w}(k)\to0$, as $k\to +\infty$. Therefore, $\gamma\left(\max_{h\geq k_0}\norm{\hat{\mathbf w}(h)}\right)\to 0$ as $k_0\to+\infty$. From \eqref{eq:diss_equilibrium}, it follows that $\mathbf x(k)\to\bar{\mathbf x}$, $\mathbf u(k)\coloneq \hat{\mathbf u}+\mathbf K(\mathbf x(k)-\hat{\mathbf x})\to\bar{\mathbf u}$, and $\mathbf y(k)\coloneq \mathbf C\mathbf x(k)\to\bar{\mathbf y}$ as $k\to +\infty$.\\
Finally, define the set 
$\bar{\mathbb X}\coloneq \prod_{i\in\mathcal M}(\delta \mathbb X_i\oplus \bar{\mathbf x}_i)$. By construction, $\bar{\mathbb X}$ is compact and contains $\bar{\mathbf x}$. Moreover, if
$x_i(k)\in\delta\mathbb X_i\oplus\bar x_i$ for all $i\in\mathcal M$, then
$\mathbf x(k)\in\bar{\mathbb X}$.\\
Suppose that $\mathbf x(0)\in\bar{\mathbb X}$. By arguments analogous to those in the proof of \cite[Proposition~2]{mayne2005robust}, it follows that $\hat {\mathbf x}(0|k)=\bar{\mathbf x}$ and $\hat {\mathbf u}(0:N-1|k)=\{\bar{\mathbf u},\dots,\bar{\mathbf u}\}$. Substitution in \eqref{eq:full_system_dynamics} implies that
\[\mathbf x(1)=f((\mathbf A+\mathbf B\mathbf K)\mathbf x(0)+\mathbf B(\bar{\mathbf u}-\mathbf K\bar{\mathbf x})).\]
Therefore, $\mathbf x(1)\in\bar{\mathbb X}$, and applying the previous argument recursively, it follows that \eqref{eq:full_system_dynamics} evolves according to
\begin{equation*}
    \mathbf x(k+1)=f((\mathbf A+\mathbf{BK})\mathbf x(k)+\mathbf B(\mathbf{\bar u}-\mathbf{K}\bar {\mathbf x})),
\end{equation*}
for all $k\geq 1$. Moreover, $\mathbf x(k)\in\bar{\mathbb X}$ for all $k\geq 1$, i.e., $\bar{\mathbb X}$ is positively invariant for \eqref{eq:full_system_dynamics}. This result, together with the asymptotic convergence of $(\mathbf x(k),\mathbf u(k),\mathbf y(k))$ towards $(\bar{\mathbf x},\bar{\mathbf u},\bar{\mathbf y})$, implies the asymptotic stability of $(\bar{\mathbf x},\bar{\mathbf u},\bar{\mathbf y})$ and completes the proof.
\hfill$\square$\smallskip\\
\textit{Proof of Proposition \ref{prop:diss_desing}.}
The proof of Proposition~\ref{prop:diss_desing} is divided into three steps, which are specified here for better clarity.
    \begin{itemize}
        \item[1.] Prove that if \eqref{eq:RPI_cstr_1} holds then \eqref{eq:subsystem_dynamics_closed_loop} is $\delta$ISS with Lyapunov function $V_i(x_i^{[a]},x_i^{[b]})=\|x_i^{[a]}-x_i^{[b]}\|^2_{P_i}$, where $x_i^{[a]},x_i^{[b]}\in\R^{n_i}$.
        \item[2.] Show that if $w_i(k)$ is given by \eqref{eq:disturbance_bound} and $u_j(k)\in\mathbb U_j$ for all $j\in\mathcal N_i$, then  \eqref{eq:RPI_cstr_2} implies that $w_i(k)\in\mathcal E(Q_{w,i})$, where $Q_{w,i}\coloneq(1+\frac{1}{\tau_i^2})P_i$.
        \item[3.] Show that items~1 and~2 imply that the set $\delta \mathbb X_i$ is RPI for $\Delta x_i(k)$.
    \end{itemize}
1. Notice that \eqref{eq:subsystem_dynamics_closed_loop} can be rewritten as
\begin{equation}\label{eq:subsystem_dynamics_closed_loop_w_tilde} 
        x_i(k+1)=f_i(A_{\mathrm K,i} x_i(k)+\tilde w_i(k)).
\end{equation}
Left and right-multiplying \eqref{eq:RPI_cstr_1}  by $\diag(P_i,P_i)$, it follows that
\[\begin{bmatrix}
    (1-\frac{1}{\gamma_i})P_i&(\tilde A_i+\tilde B_{ii}K_i)^\top P_i\\
   P_i(\tilde A_i+\tilde B_{ii}K_i)&\frac{1}{1+\tau_i^2}P_i
\end{bmatrix}\succeq0.\]
Applying the Schur complement, it follows that
\begin{equation}\label{eq:diss_condition_schur}
    \left(1-\frac{1}{\gamma_i}\right)P_i-(1+\tau_i^2)\tilde A_{\mathrm K,i}^\top P_i\tilde A_{\mathrm K,i}\succeq0.
\end{equation}
where $\tilde A_{K,i}\coloneq\tilde A_i+\tilde B_{ii}K_i$.\\
Noting that $1/\gamma_i\in(0,1)$, \eqref{eq:diss_condition_schur} implies that
\begin{equation*}
P_i-\tilde A_{K,i}^\top P_i\tilde A_{K,i}\succeq\frac{1}{\gamma_i}P_i+\tau_i^2\tilde A_{K,i}^\top P_i\tilde A_{K,i}\succ0,
\end{equation*}
which is a sufficient condition for the $\delta$ISS of \eqref{eq:subsystem_dynamics_closed_loop_w_tilde} in view of \cite[Theorem 2]{d2023incremental}. Moreover, consider two trajectories satisfying $x_i^{[a]}(k+1)=f_i(A_{\mathrm K,i}x_i^{[a]}(k)+\tilde w_i^{[a]}(k))$
and $x_i^{[b]}(k+1)=f_i(A_{\mathrm K,i}x_i^{[b]}(k)+\tilde w_i^{[b]}(k))$, then,
\begin{multline}\label{eq:diss_ly_fun}
\Delta V_i\coloneq V_i(x_i^{[a]}(k+1),x_i^{[b]}(k+1))-V_i(x_i^{[a]}(k),x_i^{[b]}(k))\\\leq -\|x_i^{[a]}(k){-}x_i^{[b]}(k)\|^2_{Q_{x,i}}{+}\|\tilde w_i^{[a]}(k){-}\tilde w_i^{[b]}(k)\|^2_{Q_{w,i}},
\end{multline}
where 
\begin{equation}\label{eq:Qxi}
Q_{x,i}\coloneq P_i-(1+\tau^2_i)\tilde A_{\mathrm K,i}^\top P_i\tilde A_{\mathrm K,i}.
\end{equation}
Note that, in view of \eqref{eq:diss_condition_schur}, $Q_{x,i}\succ0$.
\smallskip\\
2. In view of the definition of $w_i(k)$ in \eqref{eq:disturbance_bound}, a sufficient condition such that $w_i(k)\in\mathcal E(Q_{w,i})$ is 
\[\sum_{j,l\in\mathcal N_i}\left(B_{ij}(u_j(k)-\bar u_j)\right)^\top Q_{w,i}\left(B_{il}(u_l(k)-\bar u_l)\right)\leq1.\]
Noting that for any pair of vectors $v_j,v_l\in\R^{n_i}$ it holds that $v_j^\top v_l\leq\frac{1}{2}v_j^\top v_j+\frac{1}{2}v_l^\top v_l$, it follows that
\begin{align*}
&\sum_{j,l\in\mathcal N_i}\left(B_{ij}(u_j(k)-\bar u_j)\right)^\top Q_{w,i}\left(B_{il}(u_l(k)-\bar u_l)\right)\\
&\leq\sum_{j,l\in\mathcal N_i}\Big(\frac{1}{2}\left(B_{ij}(u_j(k)-\bar u_j)\right)^\top Q_{w,i}\left(B_{ij}(u_j(k)-\bar u_j)\right)\\&\quad+\frac{1}{2}\left(B_{il}(u_j(k)-\bar u_j)\right)^\top Q_{w,i}\left(B_{il}(u_l(k)-\bar u_l)\right)\Big)\\
&=2\sum_{j,l\in\mathcal N_i}\frac{1}{2}\left(B_{ij}(u_j(k)-\bar u_j)\right)^\top Q_{w,i}\left(B_{ij}(u_j(k)-\bar u_j)\right)\\
&=\nu_i\sum_{j\in\mathcal N_i}\left(B_{ij}(u_j(k)-\bar u_j)\right)^\top Q_{w,i}\left(B_{ij}(u_j(k)-\bar u_j)\right)
\end{align*}
Since $u_j(k)\in\mathbb U_j$ by assumption, it follows that $(u_j(k)-\bar u_j)\in\mathcal E(Q_{u,j}^0)$.
Therefore, a sufficient condition such that $w_i\in\mathcal E(Q_{w,i})$ is that, for all $j\in\mathcal N_i$,
\begin{equation*}
    \nu_iB_{ij}^\top Q_{w,i}B_{ij}\preceq\frac{1}{\nu_i}Q_{u,j}^0.
\end{equation*}
Recalling that $Q_{w,i}\coloneq(1{+}\frac{1}{\tau_i^2})P_i$, this condition is satisfied if
\begin{equation*}
    Q_{u,j}^0- \left(1+\frac{1}{\tau_i^2}\right)\nu_i^2B_{ij}^\top P_iB_{ij}\succeq0, \quad \forall j\in\mathcal N_i.
\end{equation*}
which, applying the Schur complement, is equivalent to
\eqref{eq:RPI_cstr_2}.\smallskip\\
3. If we consider the actual state and nominal motions, then $x_i^{[a]}(k)=x_i(k)$, $\tilde w_i^{[a]}=B_{ii}(\hat u_i(k)-K_i\hat x_i(k))+\sum_{j\in\mathcal N_i}B_{ij}\bar u_j+w_i(k)$, $x_i^{[b]}(k)=\hat x_i(k)$, and $\tilde w_i^{[b]}(k)=B_{ii}(\hat u_i(k)-K_i\hat x_i(k))+\sum_{j\in\mathcal N_i}\!B_{ij}\bar u_j$. Therefore, \eqref{eq:diss_ly_fun} implies that
\begin{equation*}
\Delta V_i\leq -\|\Delta x_i(k)\|^2_{Q_{x,i}}+\|w_i(k)\|^2_{Q_{w,i}}.
\end{equation*}
Since $w_i\in\mathcal E(Q_{w,i})$, it follows that
\begin{equation}\label{eq:dISS_lyapunov_decrease}
\Delta V_i\leq -\|\Delta x_i(k)\|^2_{Q_{x,i}}+1.
\end{equation}
Recalling \eqref{eq:Qxi}, condition \eqref{eq:diss_condition_schur} can be rewritten as
\begin{equation}\label{eq:RPI_cstr_1_schur}
    P_i/\gamma_i\preceq Q_{x,i}.
\end{equation}
Using \eqref{eq:dISS_lyapunov_decrease} and \eqref{eq:RPI_cstr_1_schur}, the set $\delta\mathbb X_i$ is RPI for $\Delta x_i(k)$ in view of \cite[Lemma 11]{ravasio2025recurrent}, completing the proof.
\hfill$\square$
\subsection{Overlapping decomposition}\label{app:overlapping_decomposition}
In this section, we show how an RNN-based plant model can be decomposed into $M$ input-coupled subplants of the form~\eqref{eq:subsystem_dynamics} by means of a fully overlapping decomposition~\cite{farina2018distributed}. Consider a plant described by
\begin{equation}\label{eq:plant_dynamics_f}
\begin{cases}
x(k{+}1)\!=\!f(Ax(k)\!+\!Bu(k))\\
y(k)=Cx(k)
\end{cases}\ 
\end{equation}
where $x \in\mathbb{R}^n$, $u\in \mathbb{R}^m$, and $y\in  \mathbb{R}^{p}$ denote the state, input, and output vectors respectively.\\ 
The input and output vectors are first partitioned into $M$ non-overlapping
subvectors, denoted as $u_i\in\R^{m_i}$ and $y_i\in\R^{p_i}$ for $i=1,\dots,M$, respectively, such that $m=\sum_{i=0}^Mm_i$ and
$p=\sum_{i=0}^Mp_i$.\\    
Accordingly, we partition matrices $B$ and $C$ as
\begin{align*}
&B=\left[B_1,\,\dots,\,B_M\right],\\
&C=\left[C_1^\top,\,\dots,\,C_M^\top\right]^\top.
\end{align*}
where $B_{ii}\in\R^{n\times m_i}$ and $C_i\in\R^{p_i\times n}$ for all $i\in\mathcal M$.\\
Hence, \eqref{eq:plant_dynamics_f} can be represented as $M$ input-coupled subsystems of the form~\eqref{eq:subsystem_dynamics} by setting $n_i=n$, $f_i(\cdot)=f(\cdot)$, $A_i=A$, $B_{ii}=B_i$, and $B_{ij}=B_j$ for all $i\in\mathcal M$ and $j\in\mathcal N_i$, with $\mathcal N_i=\mathcal M\setminus\{i\}$.
\bibliographystyle{IEEEtran}
\bibliography{biblio}

@article{ravasio2024lmi,
  title={LMI-based design of a robust model predictive controller for a class of recurrent neural networks with guaranteed properties},
  author={Ravasio, Daniele and Farina, Marcello and Ballarino, Andrea},
  journal={IEEE Control Systems Letters},
  volume={8},
  pages={1126--1131},
  year={2024},
  publisher={IEEE}
}

@article{ravasio2025recurrent,
  title={Recurrent neural network-based robust control systems with regional properties and application to {MPC} design},
  author={Ravasio, Daniele and La Bella, Alessio and Farina, Marcello and Ballarino, Andrea},
  journal={arXiv preprint arXiv:2506.20334},
  year={2025}
}

@article{riverso2013plug,
  title={Plug-and-play decentralized model predictive control for linear systems},
  author={Riverso, Stefano and Farina, Marcello and Ferrari-Trecate, Giancarlo},
  journal={IEEE Transactions on Automatic Control},
  volume={58},
  number={10},
  pages={2608--2614},
  year={2013},
  publisher={IEEE}
}

@article{mayne2005robust,
  title={Robust model predictive control of constrained linear systems with bounded disturbances},
  author={Mayne, David Q and Seron, Mar{\'\i}a M and Rakovi{\'c}, Sa{\v{s}}a V},
  journal={Automatica},
  volume={41},
  number={2},
  pages={219--224},
  year={2005},
  publisher={Elsevier}
}

@incollection{farina2018distributed,
  title={Distributed {MPC} for large-scale systems},
  author={Farina, Marcello and Scattolini, Riccardo},
  booktitle={Handbook of Model Predictive Control},
  pages={239--258},
  year={2018},
  publisher={Springer}
}

@article{cantoni2007control,
  title={Control of large-scale irrigation networks},
  author={Cantoni, Michael and Weyer, Erik and Li, Yuping and Ooi, Su Ki and Mareels, Iven and Ryan, Matthew},
  journal={Proceedings of the IEEE},
  volume={95},
  number={1},
  pages={75--91},
  year={2007},
  publisher={IEEE}
}

@inproceedings{bayer2013discrete,
  title={Discrete-time incremental ISS: A framework for robust {NMPC}},
  author={Bayer, Florian and B{\"u}rger, Mathias and Allg{\"o}wer, Frank},
  booktitle={2013 European control conference (ECC)},
  pages={2068--2073},
  year={2013},
  organization={IEEE}
}

@article{d2023incremental,
  title={An incremental input-to-state stability condition for a class of recurrent neural networks},
  author={D'Amico, William and La Bella, Alessio and Farina, Marcello},
  journal={IEEE Transactions on Automatic Control},
  volume={69},
  number={4},
  pages={2221--2236},
  year={2023},
  publisher={IEEE}
}

@article{cao2004absolute,
  title={Absolute exponential stability of recurrent neural networks with Lipschitz-continuous activation functions and time delays},
  author={Cao, Jinde and Wang, Jun},
  journal={Neural networks},
  volume={17},
  number={3},
  pages={379--390},
  year={2004},
  publisher={Elsevier}
}

@article{johansson2000,
  title={The quadruple-tank process: A multivariable laboratory process with an adjustable zero},
  author={Johansson, Karl Henrik},
  journal={IEEE Transactions on control systems technology},
  volume={8},
  number={3},
  pages={456--465},
  year={2000},
  publisher={IEEE}
}

@article{RIVERSO20142179,
title = {Plug-and-play model predictive control based on robust control invariant sets},
journal = {Automatica},
volume = {50},
number = {8},
pages = {2179-2186},
year = {2014},
issn = {0005-1098},
author = {Stefano Riverso and Marcello Farina and Giancarlo Ferrari-Trecate},
}

@ARTICLE{Riverso_pnp,
  author={Riverso, Stefano and Farina, Marcello and Ferrari-Trecate, Giancarlo},
  journal={IEEE Transactions on Automatic Control}, 
  title={Plug-and-Play Decentralized Model Predictive Control for Linear Systems}, 
  year={2013},
  volume={58},
  number={10},
  pages={2608-2614}}

@article{SCATTOLINI2009723,
title = {Architectures for distributed and hierarchical Model Predictive Control – A review},
journal = {Journal of Process Control},
volume = {19},
number = {5},
pages = {723-731},
year = {2009},
issn = {0959-1524},
author = {Riccardo Scattolini}
}

@article{BONASSI202292,
title = {On Recurrent Neural Networks for learning-based control: Recent results and ideas for future developments},
journal = {Journal of Process Control},
volume = {114},
pages = {92-104},
year = {2022},
issn = {0959-1524},
author = {Fabio Bonassi and Marcello Farina and Jing Xie and Riccardo Scattolini}
}

@misc{ravasio2026physics,
      title={Physics-informed structured learning of a class of recurrent neural networks with guaranteed properties}, 
      author={Daniele Ravasio and Claudia Sbardi and Yushan Li and Marcello Farina and Andrea Ballarino},
      year={2026},
      eprint={2603.25574} 
}

@article{FARINA20121088,
title = {Distributed predictive control: A non-cooperative algorithm with neighbor-to-neighbor communication for linear systems},
journal = {Automatica},
volume = {48},
number = {6},
pages = {1088-1096},
year = {2012},
issn = {0005-1098},
author = {Marcello Farina and Riccardo Scattolini}
}

@MISC{Siljak,
       author = {{Siljak}, Dragoslav},
        title = "{Decentralized Control of Complex Systems}",
 howpublished = {Academic Press, Inc.},
         year = 1991,
}

@article{STEWART2010460,
title = {Cooperative distributed model predictive control},
journal = {Systems {\&} Control Letters},
volume = {59},
number = {8},
pages = {460-469},
year = {2010},
issn = {0167-6911},
author = {Brett T. Stewart and Aswin N. Venkat and James B. Rawlings and Stephen J. Wright and Gabriele Pannocchia}
}

@ARTICLE{deGiuli,
  author={de Giuli, Laura Boca and La Bella, Alessio and Scattolini, Riccardo},
  journal={IEEE Transactions on Control Systems Technology}, 
  title={Physics-Informed Neural Network Modeling and Predictive Control of District Heating Systems}, 
  year={2024},
  volume={32},
  number={4},
  pages={1182-1195}}

@ARTICLE{Camponogara_distributed,
  author={Camponogara, E. and Jia, D. and Krogh, B.H. and Talukdar, S.},
  journal={IEEE Control Systems Magazine}, 
  title={Distributed model predictive control}, 
  year={2002},
  volume={22},
  number={1},
  pages={44-52},
  doi={10.1109/37.980246}}

@article{Trodden01082010,
author = {Paul Trodden and Arthur Richards},
title = {Distributed model predictive control of linear systems with persistent disturbances},
journal = {International Journal of Control},
volume = {83},
number = {8},
pages = {1653--1663},
year = {2010},
publisher = {Taylor \& Francis}}

@book{maestre2014distributed,
  title={Distributed model predictive control made easy},
  author={Maestre, Jos{\'e} M and Negenborn, Rudy R and others},
  volume={69},
  year={2014},
  publisher={Springer}
}

@book{boyd1994linear,
  title={Linear matrix inequalities in system and control theory},
  author={Boyd, Stephen and El Ghaoui, Laurent and Feron, Eric and Balakrishnan, Venkataramanan},
  year={1994},
  publisher={SIAM}
}

@article{jaeger2002adaptive,
  title={Adaptive nonlinear system identification with echo state networks},
  author={Jaeger, Herbert},
  journal={Advances in neural information processing systems},
  volume={15},
  year={2002}
}

@article{d2023data,
  title={Data-based control design for nonlinear systems with recurrent neural network-based controllers},
  author={D'Amico, William and La Bella, Alessio and Dercole, Fabio and Farina, Marcello},
  journal={IFAC-PapersOnLine},
  volume={56},
  number={2},
  pages={6235--6240},
  year={2023},
  publisher={Elsevier}
}

@article{rakovic2005invariant,
  title={Invariant approximations of the minimal robust positively invariant set},
  author={Rakovic, Sasa V and Kerrigan, Eric C and Kouramas, Konstantinos I and Mayne, David Q},
  journal={IEEE Transactions on automatic control},
  volume={50},
  number={3},
  pages={406--410},
  year={2005},
  publisher={IEEE}
}
\end{document}